\documentclass[aps,prl,twocolumn,groupedaddress]{revtex4}
\usepackage[dvips]{graphicx}
\usepackage{graphics}
\usepackage{epsfig}

\begin{document}

\title{Comment on 
%'Critical behaviour in the relaminarization of localized turbulence
%in pipe flow'.
Willis and Kerswell, PRL {\bf 98}, {014501} (2007).
}
\author{Bjorn Hof}
\author{Jerry Westerweel}
\author{Tobias M. Schneider}
\author{Bruno Eckhardt}
\affiliation{School of Physics and Astronomy, The University of Manchester, 
	Brunswick Street, Manchester M13 9PL, UK.}
\affiliation{Laboratory for Aero- and Hydrodynamics, Delft
	University of Technology, Leeghwaterstraat 21, 2628 CA Delft, The Netherlands.}
\affiliation{Fachbereich Physik, Philipps-Universit\"at Marburg, 
	Renthof 6, D-35032 Marburg, Germany}

\maketitle

In \cite{willis} Willis and Kerswell study in direct numerical simulations the
statistics of turbulent lifetimes in pipe flow. Their simulations of 
a very long pipe are of considerable interest in relation to the chaotic saddle 
picture of the transition to turbulence \cite{arfm}. They suggest that their data 
for six different Reynolds numbers support a divergence
of the lifetime near a Reynolds number of about 1870.  However, their conclusion
is not compelling: a re-analysis of their data shows that it is 
compatible with an exponentially increasing lifetime, as observed in \cite{hofnature}. 

The results in \cite{willis} are based on the turbulent lifetimes of flows which are
prepared from snapshots of a turbulent simulation at Reynolds number 1900 which is 
then integrated at a lower Reynolds number until they relaminarize or
until they reach a maximal integration time of $1000$ U/D. From 40 to 60 such initial conditions
the probability $P(T)$ to remain turbulent for a time $T$ at least is obtained and compared to 
an exponential variation, $P(T)\sim\exp(-T \ln 2/\tau_h)$, with $\tau_h(Re)$ the characteristic
lifetime.
The problematic part is that this exponential distribution is expected to occur for long times
only \cite{kadanoff}, and that the choice of preparation of initial conditions introduces
transient elements into $P(t)$ of unpredictable duration. This is evident from a magnification
of their Fig.~3, shown here as Fig.~1. $P(T)$ for the lowest Reynolds number 
$Re=1580$ clearly is constant up to about $38$ U/D, drops off slowly until $58$ U/D and
then falls of more steeply for longer times. An exponential fit to the tail for times $T>58$ 
gives a lifetime of $\tau_h\approx 4.9$, much shorter than the value that can be read off from 
their Fig.~2, $\tau_{h,WK}\approx 14.5$.  The lifetime statistics at $Re=1700$, $1740$, 
$1780$, and $1820$ also show several regions with different behaviour, and again only the 
tails should be fitted. The data at $Re=1860$ is inconclusive, since it covers a small range
of $P(T)$ only.

\begin{figure}
   \epsfig{figure=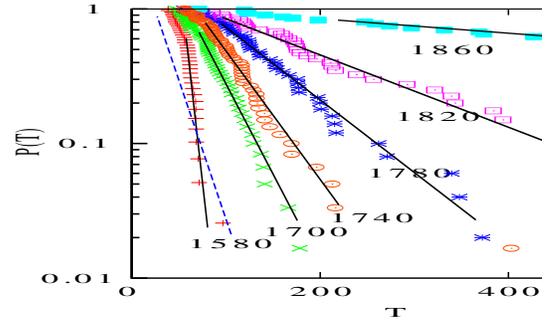, angle=0, width=70mm}\\[-1em]
   \caption[]{Turbulent lifetimes $P(T)$ for pipe flow. Shown is a magnification of 
Fig.~3 from \cite{willis} (symbols and Reynolds numbers as used there) together with
straight lines indicating the fit to an asymptotic exponential (straight line). 
The dashed line for $Re=1580$ has the slope they use in their analysis. 
% from Fig.~4 of \cite{willis}.
\\[-2em]
}
\end{figure}

\begin{figure}
   \epsfig{figure=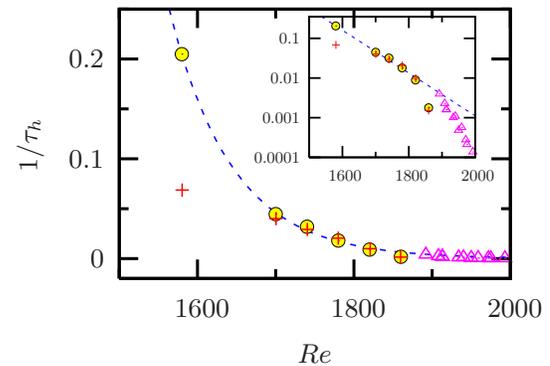, angle=0, width=70mm}\\[-1em]
   \caption[]{Variation of the half lifetime with $Re$. 
The red crosses are the data of \cite{willis},
the circles the ones read off from the fits shown in Fig.~1 and the margenta triangles are the
data of \cite{hofnature}. The dashed line is a fit to an exponential variation. The inset shows the
same data on a logarithmic scale.\\[-1em]
}
\end{figure}

The re-analysis of the slopes has consequences for the variations of 
lifetimes with Reynolds number, as shown in Fig.~2. 
The reduced lifetimes at $Re=1580$ give a higher value for $1/\tau_h$. 
The lifetimes extracted from Fig.~1 lie close to the experimental data
of \cite{hofnature}, and are compatible with an exponential
variation with
Reynolds number. 
A logarithmic representation shows that the last data point lies somewhat lower, and that
there are small differences in the slopes. 
Unfortunately, the simulations stop where the 
experiments begin, so it is difficult to see whether the observed differences point to 
a transition in the character of the turbulent dynamics. 
We conclude that the reanalysis of the data shows that they are compatible with an 
exponential increase of the lifetimes, as observed in \cite{hofnature}.\\[-2em]


\begin{thebibliography}{4}
\expandafter\ifx\csname natexlab\endcsname\relax\def\natexlab#1{#1}\fi
\expandafter\ifx\csname bibnamefont\endcsname\relax
  \def\bibnamefont#1{#1}\fi
\expandafter\ifx\csname bibfnamefont\endcsname\relax
  \def\bibfnamefont#1{#1}\fi
\expandafter\ifx\csname citenamefont\endcsname\relax
  \def\citenamefont#1{#1}\fi
\expandafter\ifx\csname url\endcsname\relax
  \def\url#1{\texttt{#1}}\fi
\expandafter\ifx\csname urlprefix\endcsname\relax\def\urlprefix{URL }\fi
\providecommand{\bibinfo}[2]{#2}
\providecommand{\eprint}[2][]{\url{#2}}
\bibitem[{\citenamefont{Willis and Kerswell}(2007)}]{willis}
\bibinfo{author}{\bibfnamefont{A.~P.} \bibnamefont{Willis}} \bibnamefont{and}
  \bibinfo{author}{\bibfnamefont{R.~R.} \bibnamefont{Kerswell}},
  \bibinfo{journal}{Phys. \ Rev.\ Lett.} \textbf{\bibinfo{volume}{98}},
  \bibinfo{pages}{014501} (\bibinfo{year}{2007}).

\bibitem[{\citenamefont{Eckhardt et~al.}(2007)\citenamefont{Eckhardt,
  Schneider, Hof, and Westerweel}}]{arfm}
\bibinfo{author}{\bibfnamefont{B.}~\bibnamefont{Eckhardt}},
  \bibinfo{author}{\bibfnamefont{T.~M.} \bibnamefont{Schneider}},
  \bibinfo{author}{\bibfnamefont{B.}~\bibnamefont{Hof}}, \bibnamefont{and}
  \bibinfo{author}{\bibfnamefont{J.}~\bibnamefont{Westerweel}},
  \bibinfo{journal}{Annual Review Fluid Mechanics}
  \textbf{\bibinfo{volume}{39}}, \bibinfo{pages}{447} (\bibinfo{year}{2007}).

\bibitem[{\citenamefont{Hof et~al.}(2006)\citenamefont{Hof, Westerweel,
  Schneider, and Eckhardt}}]{hofnature}
\bibinfo{author}{\bibfnamefont{B.}~\bibnamefont{Hof}},
  \bibinfo{author}{\bibfnamefont{J.}~\bibnamefont{Westerweel}},
  \bibinfo{author}{\bibfnamefont{T.~M.} \bibnamefont{Schneider}},
  \bibnamefont{and} \bibinfo{author}{\bibfnamefont{B.}~\bibnamefont{Eckhardt}},
  \bibinfo{journal}{Nature} \textbf{\bibinfo{volume}{443}}, \bibinfo{pages}{55}
  (\bibinfo{year}{2006}).

\bibitem[{\citenamefont{L.P.Kadanoff and Tang}(1984)}]{kadanoff}
\bibinfo{author}{\bibnamefont{L.P.Kadanoff}} \bibnamefont{and}
  \bibinfo{author}{\bibfnamefont{C.}~\bibnamefont{Tang}},
%  \bibinfo{journal}{Proc. Natl Acad. Sci. USA} \textbf{\bibinfo{volume}{81}},
  \bibinfo{journal}{PNAS} \textbf{\bibinfo{volume}{81}},
  \bibinfo{pages}{1276} (\bibinfo{year}{1984}).

\end{thebibliography}
\end{document}